\documentclass[aps,pra,twocolumn,superscriptaddress,amsmath,showpacs,nofootinbib]{revtex4}
\usepackage{amsmath} \usepackage{graphicx} \usepackage{amssymb}
\usepackage{epsfig} \usepackage{amsfonts}

\usepackage{color}

 \def\fig#1{{#1}}

\newcommand{\etal}{{\em et al.}}
\newcommand{\tr}{\mathrm{Tr}}
\newcommand{\Tr}{\mathrm{Tr}}

\newcommand{\I}{i}
\newcommand{\kd}[1]{\delta_{#1}}
\newcommand{\e}[1]{\varepsilon_{#1}}
\newcommand{\kron}{\otimes}

\begin{document}

\title{Direct method for measuring purity,
superfidelity, and subfidelity\\ of photonic two-qubit mixed
states}

\author{Karol Bartkiewicz}\email{bartkiewicz@jointlab.upol.cz}

\affiliation{RCPTM, Joint Laboratory of Optics of Palack\'y
University and Institute of Physics of Academy of Sciences of the
Czech Republic, 17. listopadu 12, 772 07 Olomouc, Czech Republic
}
\affiliation{Faculty of Physics, Adam Mickiewicz University,
PL-61-614 Pozna\'n, Poland}

\author{Karel Lemr}

\affiliation{RCPTM, Joint Laboratory of Optics of Palack\'y
University and Institute of Physics of Academy of Sciences of the
Czech Republic, 17. listopadu 12, 772 07 Olomouc, Czech Republic }

\author{Adam Miranowicz}

\affiliation{Faculty of Physics, Adam Mickiewicz University,
PL-61-614 Pozna\'n, Poland}

\begin{abstract}

The Uhlmann-Jozsa fidelity (or, equivalently, the Bures distance)
is a basic concept of quantum communication and quantum
information, which however is very difficult to measure efficiently
without recourse to quantum tomography. Here we propose
a direct experimental method to estimate the
fidelity between two unknown two-qubit mixed states via the
measurement of the upper and lower bounds of the fidelity, which
are referred to as the superfidelity and subfidelity,
respectively. Our method enables a direct measurement of the
first- and second-order overlaps between two arbitrary two-qubit
states. In particular, the method can be applied to measure the
purity (or linear entropy) of a single two-qubit mixed state in a
direct experiment. We also propose and critically compare several
experimental strategies for measuring the sub- and superfidelities
of polarization states of photons in various linear-optical
setups.
\end{abstract}


\maketitle

\section{Introduction}

Fidelity plays a fundamental role in classical~\cite{CoverBook}
and quantum~\cite{NielsenBook,BengtssonBook} communication
theories as a quantitative measure of the accuracy of imperfect
transmission of signals through a communication channel. Fidelity
has also other basic applications in quantum information, quantum
optics, and even condensed-matter physics.

The most popular definition of fidelity between two mixed quantum
states $\rho_1$ and $\rho_2$ (corresponding to, e.g., the input
and output states of a communication channel) was given by
Uhlmann~\cite{Uhlmann76} and Jozsa~\cite{Jozsa94} as
\begin{equation}
F(\rho_1,\rho_2)\equiv \Big[ {\rm
Tr}\Big(\sqrt{\sqrt{\rho_1}\rho_2\sqrt{\rho_1}}\Big) \Big]^2,
\label{fidelity}
\end{equation}
which is also referred to as the Uhlmann transition
probability~\cite{Uhlmann76}. To avoid confusion, we note that the
root fidelity $\sqrt{F}$ is sometimes referred to as the fidelity
(see, e.g., Ref.~\cite{NielsenBook}). The fidelity vanishes for
orthogonal states and is equal to 1 for identical states. If one
of the states is pure, say $\rho_1=|\psi_1\rangle\langle\psi_1|$,
then the fidelity simplifies to $F=\langle \psi_1|\rho_2 |\psi_1
\rangle$. The fidelity has a few important and useful properties
including~\cite{Uhlmann76,Jozsa94,Mendonca08,Miszczak09} (a)
bounds $0\le F(\rho_1,\rho_2)\le 1$, (b) symmetry
$F(\rho_1,\rho_2)= F(\rho_2,\rho_1)$, (c) unitary invariance
[i.e., $F(\rho_1,\rho_2)=F(U\rho_1U^{\dagger},U\rho_2U^{\dagger})$
for an arbitrary unitary operator $U$],  (d) multiplicativity
[e.g., $F(\rho_1 \otimes \rho_2,\rho_3 \otimes \rho_4)=
F(\rho_1,\rho_3) F(\rho_2,\rho_4)$], (e) concavity, and (f)
joint concavity.

A physical interpretation of the fidelity as a measure of
distinguishability can be given as
follows~\cite{Uhlmann76,Jozsa94}:  In a decoherence
scenario, based on purifications instead of collapses of states,
an arbitrary mixed state $\rho_S$ can be given by a \emph{pure}
state $|\psi_{SE}\rangle$ of a subsystem $S$ entangled with some
larger system (environment) $E$, which is reduced to $S$, i.e.,
$\rho_S=\tr_E(|\psi_{SE}\rangle\langle\psi_{SE}|)$. Then the
fidelity corresponds to the maximum taken over all such
purifications $|\psi_n\rangle\equiv|\psi_{SE}^{(n)}\rangle$ of
states $\rho_{n}\equiv \rho^{(n)}_S$ for $n=1,2$, i.e.,
$F(\rho_1,\rho_2)=\max|\langle\psi_1|\psi_2\rangle|^2$.

The fidelity is simply related to the Bures metric~\cite{Bures69}
(the Helstrom metric~\cite{Helstrom67}) as $D_B(\rho_1,\rho_2)^2 =
2[1-\sqrt{F(\rho_1,\rho_2)}]$, which can be considered a quantum
generalization of the Fisher information metric. The Bures metric
can be used to quantify quantum
entanglement~\cite{Vedral97,Marian08},
nonclassicality~\cite{Marian02}, and
polarization~\cite{polarization}. The Braunstein-Caves
distinguishability metric~\cite{Braunstein94}, defined via the
Bures metric, is a useful tool of quantum estimation theory.
Entanglement measures based on the Bures metric are also useful for
identifying and characterizing quantum phase transitions, e.g., as
indicators of their criticality~\cite{phase_transitions}.

An important question arises as to how to measure the fidelity
$F(\rho_1,\rho_2)$ (or its bounds) between two mixed states. Obviously, one
can apply a method for quantum state tomography for the complete
reconstruction of the states $\rho_1$ and $\rho_2$. Then, with this knowledge, the
fidelity can be calculated explicitly. However, this approach is
extremely inefficient as it requires measuring redundant information
to finally determine just a single value of the fidelity. Note that the
fidelity, given by Eq.~(\ref{fidelity}), between two \emph{mixed}
states is difficult not only to measure directly but even to
calculate analytically. We note that analytical formulas for
the fidelity are known only for a few types of states including
single-qubit states~\cite{BengtssonBook} and multimode Gaussian
states~\cite{Marian12}.

In this article we show how to directly measure the super- and
subfidelities which are, respectively, the upper and lower bounds on the
fidelity $F(\rho_1,\rho_2)$ between two arbitrary two-qubit mixed
states $\rho_1$ and $\rho_2$~\cite{Miszczak09,Zhou12}.
Moreover, we  describe here an experimental method for
measuring the purity $\chi(\rho)$, which is a degree of
information about the preparation of a quantum state $\rho$ as
defined by the trace norm of $\rho$:
\begin{equation}
\chi(\rho)=||\rho\rho^\dagger||= ||\rho^2|| = \tr(\rho^2).
\end{equation}
The purity ranges from $\chi=1/d$ for  completely mixed states
$\rho=I/d$ of dimension $d$ to $\chi=1$ for pure states. By
recalling deep mathematical and physical similarities between
the classical optical polarizations and qubits, one can
conclude that the classical degree of polarization and the quantum
purity of a qubit are close analogs~\cite{Gamel12,Bartkiewicz10}.

On the other hand, the lack of information about the preparation
of a given state, which is referred to as the mixedness, can be
quantified by entropic measures, such as the von Neumann and
Bastiaans-Tsallis entropies, including the linear entropy $S_L = 1
-\chi$, which is a linear approximation to the von Neumann
entropy. The linear entropy can be considered a measure of quantum
entanglement of a bipartite pure state if one of the subsystems is
traced out. For two-qubit mixed states, the relation between the
linear entropy and (i) the von Neumann entropy, (ii) the
entanglement of formation (which is a canonical measure of
entanglement), and (iii) the violations of the
Clauser-Horne-Shimony-Holt (CHSH) inequality were studied in,
e.g., Refs.~\cite{Munro01,Wei03,Peters04}. A comparative analysis
of the mixedness (as measured by the linear and von Neumann
entropies) and quantum noise (as described by the squeezing and  Fano
factors) was given in a dynamical scenario in, e.g.,
Ref.~\cite{Bajer04}.

It is also important to note that the purity $\chi$ is a special
case of the first-order overlap function
\begin{equation}
O(\rho_1,\rho_2) = \tr(\rho_1\rho_2), \label{overlap1}
\end{equation}
which is equivalent to the purity if  $\rho_1=\rho_2$.

General methods to measure any polynomial function of a
density matrix were described by Ekert \etal~ \cite{Ekert02}
and Brun~\cite{Brun04}. While these approaches would also
enable a direct measurement of the purity in a special case,
it is not clear whether they enable efficient experimental
implementations (except, e.g., the single-qubit purity
measurements~\cite{Adamson08}). Indeed, as mentioned in
Ref.~\cite{Brun04}: ``This proof of principle is very far
from being a proof that such a measurement is practical.''
These general approaches consist of using various two-qubit
and single-qubit gates such as the controlled-SWAP or
controlled-NOT and Hadamard gates. Note that the success
probability of the discrete-variable controlled-NOT and controlled-SWAP
gates with linear optics is limited, e.g., it is usually
equal to 1/9 assuming no additional ancillae and feedforward
(see Ref.~\cite{Bartkowiak10} and references therein). Thus,
a setup based on these approaches would be less efficient
than specifically dedicated setups, such as the one proposed
in this paper.

In this article we show how the first-order overlap
$O(\rho_1,\rho_2)$ [and, thus, the purity $\chi(\rho)$] can be
directly  measured for arbitrary single- and
two-qubit mixed states in a linear-optical experiment. We note
that some experimental works on directly measuring the purity have
already been reported by Du \etal~ \cite{Du04} in liquid-state NMR
systems of three spins-1/2 and, by using quantum polarization
states as qubits, by Bovino \etal~\cite{Bovino05} in a four-photon
system, and by Adamson \etal~ \cite{Adamson08} (based on the method
proposed by Brun~\cite{Brun04}) in two- and three-photon systems.
We note  that our method requires a smaller number of
detectors in comparison to, e.g., the method of Bovino
\etal~ \cite{Bovino05}. The problem of measuring the
purity of a quantum state (and the overlap between two quantum
states) within a `minimal' model was theoretically studied in
Ref.~\cite{Tanaka13}. The first theoretical proposals for measuring
the sub- and superfidelities were described in detail by Miszczak
\etal~\cite{Miszczak09} by applying the methods of Ekert
\etal~\cite{Ekert02} and Bovino \etal~\cite{Bovino05},
respectively.

Our proposals for experiments on the first- and second-order
overlaps are inspired by the method for the measurement of
nonclassical correlations described in detail in
Ref.~\cite{Bartkiewicz13}, which can also be used for measuring,
e.g., the degree of the CHSH inequality
violation~\cite{Bartkiewicz13b}.

This article is organized as follows: In
Sec.~\ref{sec:preliminaries}, we recall some basic definitions of
the sub- and superfidelities. Moreover, we perform a Monte Carlo
simulation of the fidelity as a weighted mean of these fidelity
bounds. In Sec.~\ref{sec:setup1}, we describe direct 
methods for measuring the first-order overlap, purity, and
superfidelity. Experimental considerations are presented in
Sec.~\ref{sec:experiment}. In Sec.~\ref{sec:setup2}, we propose
direct and experimentally-friendly methods for measuring the
second-order overlap and the subfidelity. We conclude in
Sec.~\ref{sec:conclusion}.

\section{Subfidelity, superfidelity and the estimation of
fidelity}\label{sec:preliminaries}

The density matrix of a single qubit in the Bloch representation can
be written compactly as
\begin{equation}
\rho  = \frac{1}{2}R_{m0}\,\sigma _{m}, \label{rho1}
\end{equation}
by using the Einstein summation convention. The elements
$R_{m0}=\tr(\rho\sigma_m)$ of the Bloch vector are defined by the Pauli matrices
$\sigma_m$ for $m=0,1,2,3,$ where $\sigma_0 = I$ is the identity
operator. Furthermore,  we study  two-qubit (quartit) systems
described by density matrices $\rho$, which are expressed in the
standard Bloch representation as
\begin{equation}
\rho  = \frac{1}{4}R_{mn}\,\sigma _{m}\otimes \sigma _{n}
\label{rho}
\end{equation}
in terms of the correlation-matrix elements
$R_{mn}=\tr(\rho\sigma_m\otimes \sigma_n)$, with $m,n=0,...,3$. We
can see that a single-qubit density matrix can be obtained after
tracing out the second qubit from the two-qubit density matrix.
Thus, in this article we focus on quartits, bearing in mind that
the qubit case can be obtained simply after taking one of the
qubits out of the picture.

Let us denote compactly $F(\rho_1,\rho_2) = [\tr(A)]^2,$ where
$A=\sqrt{\sqrt{\rho_1}\rho_2\sqrt{\rho_1}}$. For single-qubit
states, $A$ is a $2\times 2$ matrix, which satisfies the
characteristic equation
\begin{equation}
A^2 - A\tr(A) + I \mathrm{det}(A) = 0;
\end{equation}
thus
\begin{equation}
F(\rho_1,\rho_2) =O(\rho_1,\rho_2) +
\sqrt{[1-O(\rho_1,\rho_1)][1-O(\rho_2,\rho_2)]},
\end{equation}
which can be directly measured by our proposed method.

For two-qubit (and higher-dimensional) density matrices the
situation is quite different. Nevertheless, we can use the upper
and lower bounds on the fidelity, given by~\cite{Miszczak09}:
\begin{eqnarray}
E(\rho_1,\rho_2)\le F(\rho_1,\rho_2) \le G(\rho_1,\rho_2),
\end{eqnarray}
where
\begin{eqnarray}
E(\rho_1,\rho_2)= \tr(\rho_1\rho_2) +
\sqrt{2[\tr(\rho_1\rho_2)]^2-2\tr[(\rho_1\rho_2)^2]},\hspace{4mm}
\label{E}\\
G(\rho_1,\rho_2)= \tr(\rho_1\rho_2) +
\sqrt{[1-\tr(\rho_1^2)][1-\mathrm{Tr(}\rho_2^2)]},\hspace{1cm}
\label{G}
\end{eqnarray}
which are referred to as the subfidelity and superfidelity,
respectively. These formulas can be rewritten as
\begin{eqnarray}
G(\rho_1,\rho_2)&=& O(\rho_1,\rho_2) + \sqrt{S_L(\rho_1)S_L(\rho_2)}, \\
E(\rho_1,\rho_2)&=& O(\rho_1,\rho_2) + \sqrt{2[
O^2(\rho_1,\rho_2)-O'(\rho_1,\rho_2)]}.\quad
\end{eqnarray}
Thus, to measure these bounds directly, we need to measure the
first-order overlap $O(\rho_1,\rho_2)$, given by
Eq.~(\ref{overlap1}), and the second-order overlap
\begin{eqnarray}
  O'(\rho_1,\rho_2) &=& \tr(\rho_1\rho_2\rho_1\rho_2),
\label{overlap2}
\end{eqnarray}
together with the linear entropies (purities),
$S_L(\rho_n)=1-O(\rho_n,\rho_n)$ (for $n=1,2$), which are the
special cases of $O(\rho_1,\rho_2)$. Let us note that if (at
least) one of the states $\rho_1,\rho_2$ is pure, then the
superfidelity is equal to the fidelity and is given only by the
first term, i.e., the overlap $O(\rho_1,\rho_2)$. Thus, after
finding that either $\rho_1$ or $\rho_2$ is pure, the fidelity can be
found by measuring the overlap $O(\rho_1,\rho_2)$ only.

\begin{figure}
\fig{
\includegraphics[width=4.25cm]{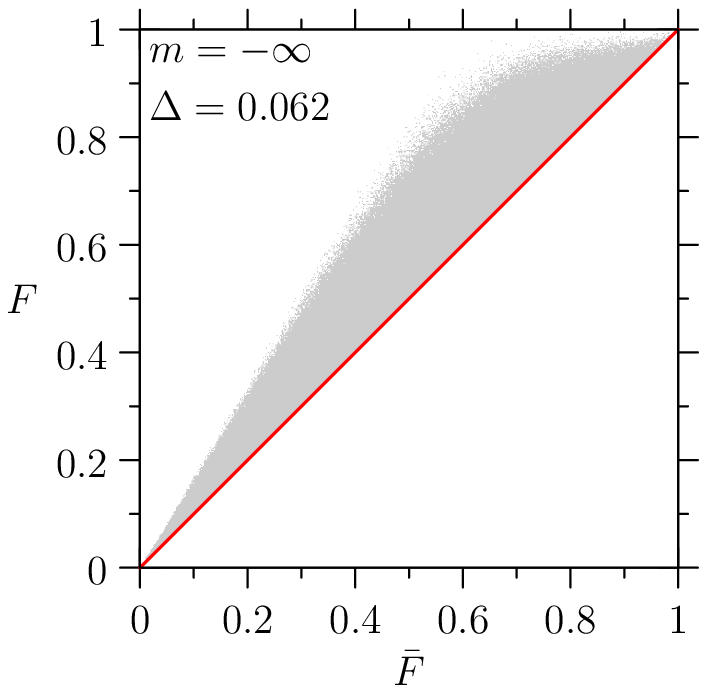}
\includegraphics[width=4.25cm]{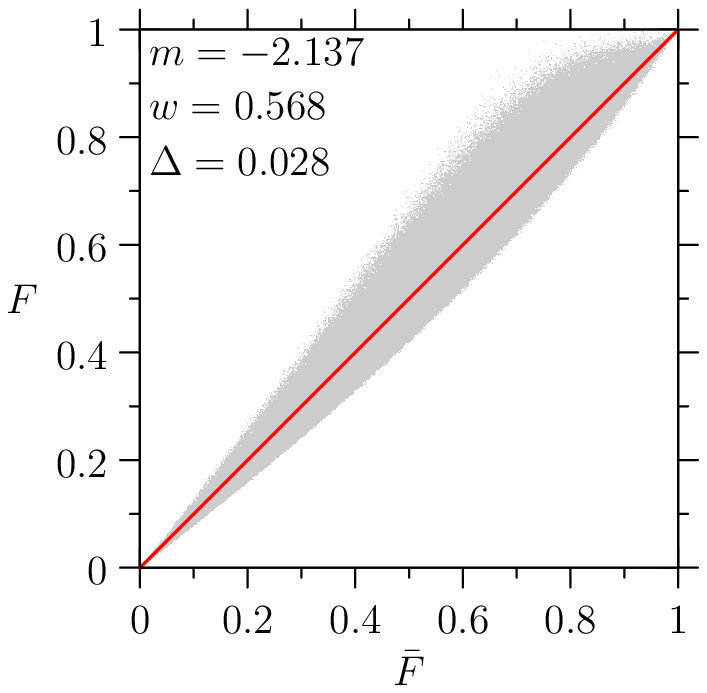}
\includegraphics[width=4.25cm]{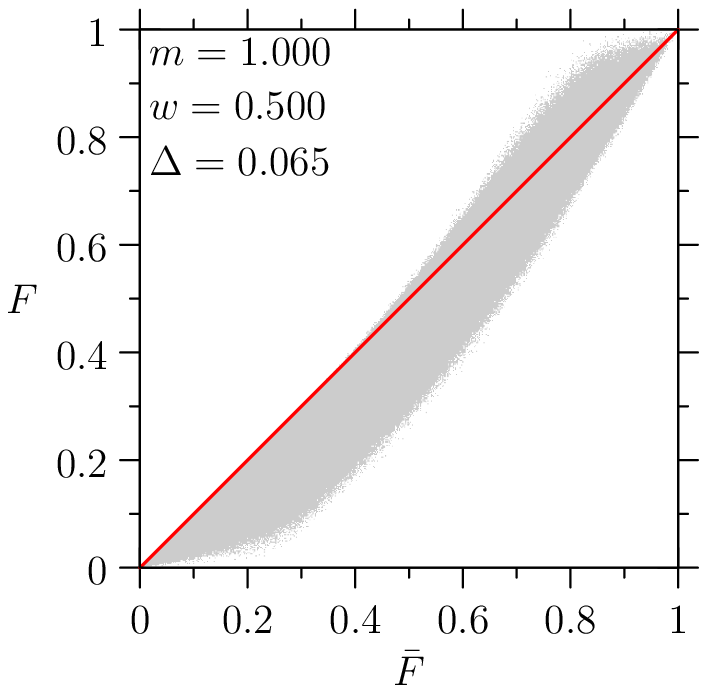}
\includegraphics[width=4.25cm]{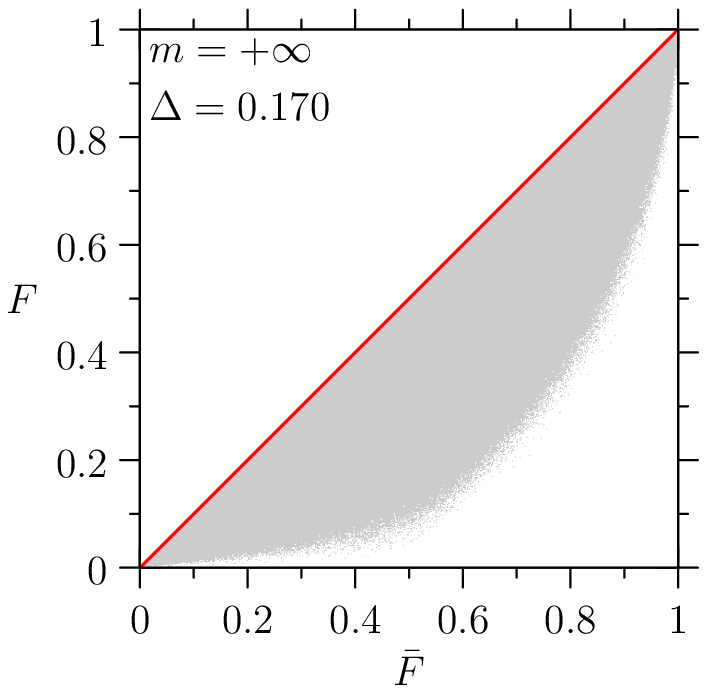}
}

\caption{\label{fig:0} (Color online) Fidelity $F(\rho_1,\rho_2)$
versus generalized mean $\bar{F}(\rho_1,\rho_2)$, corresponding to
(a) subfidelity $E(\rho_1,\rho_2)$ for $m=-\infty$, (b) optimized
mean for $m=-2.137$, (c) arithmetic mean for $m=1$, and (d)
superfidelity $G(\rho_1,\rho_2)$ for $m=+\infty$, as calculated in
our Monte-Carlo simulation for $10^7$ pairs of two-qubit states.
The estimation error is given by $\Delta=\sqrt{\langle(\bar{F} -
F)^2\rangle}$. Note that if the fidelity bounds $E$ and $G$ were
tighter, then the area covered by the simulation outcomes would be
smaller converging to the $G=E$ line. }
\end{figure}

In any other case the fidelity $F$ can be estimated as an average
of the subfidelity $E$ and superfidelity $G$ with some certain error.
The error can be minimized in two ways: (i) by finding tighter
measurable bounds on the fidelity or (ii) by using the best expression for
the mean value (arithmetic, harmonic, geometric, etc.). In our article
we  focus only on this second aspect and numerically optimize the
generalized mean (power mean) defined as
\begin{equation}
\bar{F}(\rho_1,\rho_2) =
\big[wE^m(\rho_1,\rho_2)+(1-w)G^m(\rho_1,\rho_2)\big]^{\frac{1}{m}},\label{mean}
\end{equation}
which for balanced weights $w=1-w=1/2$, and in special cases for
$m=-1,0,1$, becomes the harmonic, geometric, and arithmetic
mean, while for $m=-\infty,\infty$ it reduces to $\bar{F} =
E,G$, respectively. Thus, the generalized mean is bounded from
below by $E$ and from above by $G$, i.e.,  for
$m\in(-\infty,\infty)$
\begin{equation}
E(\rho_1,\rho_2)\le \bar{F}(\rho_1,\rho_2) \le G(\rho_1,\rho_2).
\end{equation}
From our Monte Carlo simulation for $10^7$ pairs of random
two-qubit states (as shown in Fig.~\ref{fig:0}) we found with the
method of least squares that the optimal $m = -2.13$ and $w =
0.568$ providing the smallest estimation error
$\Delta=\sqrt{\langle(\bar{F} - F)^2\rangle} = 0.0278$, which is
an improvement in comparison to the arithmetic mean providing
$\Delta=0.0652$. Our method of calculating means is not the most
general one (e.g., one can use a generalized $f$ mean). However,
one must remember that true fidelity values are to be found
between $E$ and $G$, so the maximal estimation error is greater
than the error average.

We focus on the  analysis of nonlinear properties of two-qubit states because they play an important role in
quantum protocols exploiting quantum correlations. Thus,
establishing methods of testing  various properties of these states  is well motivated. This is especially
important for photonic qubits since photons are
typical carriers of quantum information used in quantum
communication protocols.

\section{Efficient measurements of first-order overlap,
purity, and superfidelity} \label{sec:setup1}

\begin{figure}
\fig{\includegraphics[width=6cm]{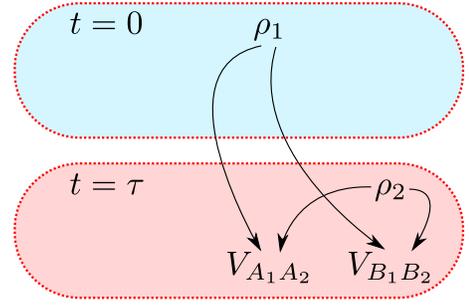}}
\caption{\label{fig:1} (Color online)  Method for a single
iteration of the measurement of the overlap
$\Tr(\rho_1\rho_2)$ (and, in a special case, the purity
$\chi$ and linear entropy $S_L=1-\chi$) between any two-qubit
states $\rho_1$ and $\rho_2$ produced at constant time
intervals $\tau$. The subsystem of the first (second) qubit
is called $A$ ($B$). The measurement should be repeated until
some large enough number  $K$ of values is accumulated. The
delay is implemented at times $t=2k\tau$, where
$k=0,1,2,...,K-1$ and $2K\tau$ is the duration of the
measurement. The delay can be implemented using fast
switching between, e.g., two paths of the optical-length
difference corresponding to delay $\tau$. This method works
also if the states $\rho_1$ and $\rho_2$ are qubits given
that we removed the measurement $V_{B_1B_2}$ or
$V_{A_1,A_2}$. }
\end{figure}

The purity $\chi$ can be observed directly if we assume that we
have access to two copies of the two-qubit system. The first-order overlap (or
the purity in the special case for $\rho_1 = \rho_2$)
 can be calculated directly as
\begin{equation}\label{eq:gamma1}
O(\rho_1,\rho_2) = \frac{1}{16}R^{(1)}_{mn}R^{(2)}_{kl}\tr
[(\sigma_m\sigma_k)\otimes(\sigma_k\sigma_l)],
\end{equation}
where $R^{(k)}$ are the correlation matrices of $\rho_k$ for
$k=1,2$, as defined below Eq.~(\ref{rho}). The multiplication of
the Pauli matrices is given as
\begin{equation}
\sigma_a\sigma_b = \I\varepsilon_{abc}\sigma_c
+\delta_{ab}\sigma_0,
\end{equation}
where $\I$ is the imaginary unit, $\delta_{ab}$ is the Kronecker
$\delta$, and $\varepsilon_{abc}$ is the Levi-Civita symbol, which
is $\varepsilon_{abc}=0$ if $a\times b \times c=0$ or at least
two indices are equal. Note that $\sigma_n$ matrices are traceless
except for $n = 0$; thus
\begin{equation}
\tr(\sigma_a\sigma_b) = 2\delta_{ab}.
\end{equation}
Hence, we can rewrite  Eq.~(\ref{eq:gamma1}) as
\begin{equation}
O(\rho_1,\rho_2) = \frac{1}{4} R^{(1)}_{mn}R^{(2)}_{mn}.
\end{equation}
We can express  $R^{(k)}_{mn}$ as expectation values of  the Pauli
matrices:
\begin{eqnarray}
R^{(1)}_{mn} &=& \tr[(\sigma_m\otimes\sigma_n)\rho_1],\\
R^{(2)}_{mn} &=& \tr[(\sigma_m\otimes\sigma_n)\rho_2],
\end{eqnarray}
hence
\begin{eqnarray}
O(\rho_1,\rho_2)  &=& \frac{1}{4}\tr
[(\sigma_m\otimes\sigma_n\otimes\sigma_m\otimes\sigma_n)\nonumber
(\rho_1\otimes\rho_2)]\\
&=& \frac{1}{4}\tr
[(\sigma_m\otimes\sigma_m)\otimes(\sigma_n\otimes\sigma_n)\nonumber
(\rho_1\otimes\rho_2)']\\
&=& \frac{1}{4}\tr [(V_{A_1A_2}\otimes V_{B_1B_2})'
(\rho_1\otimes\rho_2)] ,
\end{eqnarray}
where $V=\sigma_m\otimes\sigma_m = 2I^{\otimes2} - 4 |\Psi^-\rangle\langle\Psi^-|$,
with $|\Psi^-\rangle$ denoting the singlet state, and
$(\rho_1\otimes\rho_2)'=S_{A_2B_1}(\rho_1\otimes\rho_2)S_{A_2B_1}$.
The  self-adjoint transformation $S_{A_2B_1}=I\otimes S\otimes I$
is  the operation swapping modes $A_2$ and $B_1$, which is given in
terms of  the SWAP operator
\begin{equation}
S =\left( \begin{array}{cccc}
1 & 0 & 0 & 0\\
0 & 0 & 1 & 0\\
0 & 1 & 0 & 0\\
0 & 0 & 0 & 1
\end{array}\right).
\end{equation}
We can now introduce the Hermitian overlap operator
\begin{equation}
\Gamma = S_{A_2B_1}V_{A_1A_2}V_{B_1B_2}S_{A_2B_1},\label{Gamma}
\end{equation}
which, if measured on $\rho_1\otimes\rho_2$, provides the value of
the overlap $O(\rho_1,\rho_2)$ and, in the special case, the purity if
$\rho_1=\rho_2$. Hence, the overlap is a real observable measured
on a system consisting of two copies of the investigated state
$\rho$. Let us note that measuring the purity is,  thus,
equivalent to measuring a product of $V$ operators, which was
shown in Ref.~\cite{Bartkiewicz13} to be experimentally accessible
using linear optics.  If the two-qubit state is produced at
frequency $1/\tau$,  we delay every second state $\rho$ and we
can measure the product of $V$ operators directly as shown in
Fig.~\ref{fig:1}.

\begin{figure}
\vspace*{5mm} \fig{\includegraphics[width=8cm]{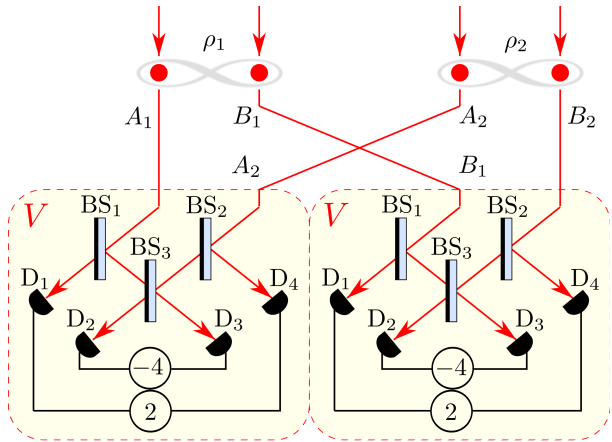}}
\caption{\label{fig:purity} (Color online) Setup implementing
a direct  measurement of the  first-order overlap
$O(\rho_1,\rho_2)=\tr(\rho_1\rho_2)$ between arbitrary
two-qubit mixed states $\rho_1$ and $\rho_2$. The setup
consists of  two $V$ blocks as in Ref.~\cite{Bartkiewicz13}.
Due to the probabilistic nature of the path taken by photons
after the BS interaction  the setup can provide conclusive
results in at most $1/4$ of the cases when $\rho_1$ and
$\rho_2$ are delivered (as discussed in Fig.~\ref{fig:1}).
This maximum performance is achieved when the photons are
deterministically antibunched (i.e., always projected onto
the singlet state) if both of them impinge on BS$_3$. Note
that the  BSs are asymmetric, i.e., the photon is phase
shifted by $\pi$ only when reflected from the black-shaded
side of the BSs. If $\rho_1$ and $\rho_2$ are single photons
then only one $V$ block is required and the measurement setup
corresponds to the one used in Ref.~\cite{Hendrych03}.}
\end{figure}

The setup shown in Fig.~\ref{fig:purity} can be used in a direct
measurement of the overlap  (thus, also of  the linear entropy)
of a two-qubit state, since $O(\rho_1,\rho_2) =
\langle\Gamma\rangle=\tr[(V_{A_1A_2}\otimes
V_{B_1B_2})(\rho_1\otimes\rho_2)']/4$.  The possible outcomes for a
single measurement instance  are $a_k =-8,0,4,16$, which are the
products of two outcomes of the coincidence detections in the left and
right  arms of the setup  in Fig.~\ref{fig:purity} marked as
$-4,0,2$. The outcome $2$ is assigned to  the coincidence
detection in the outermost detectors of  the $V$ blocks, $-4$ for
the coincidence detection in the middle detectors, and $0$ if neither
of the two coincidences has been detected.  Useful values of
$a_k$ appear for  $K_0=\eta^4 K/4$ of the cases when the states
$\rho_1$ and $\rho_2$ are delivered, assuming that $\eta$ is the
quantum efficiency of the detectors. The obtained expectation
value of the overlap reads
\begin{equation}
O(\rho_1,\rho_2)= \langle \Gamma \rangle  \approx
\frac{1}{4K_0}\sum_{k=1}^K a_k,
\end{equation}
where
\begin{equation}
K_0 = \sum_{k=1}^{K}\delta_{a_k,4}
\end{equation}
and $\delta_{a_k,4}$ is the Kronecker $\delta$. As for any
measurement, equality is reached  in the limit of $K_0\to\infty$.
Thus, we  have demonstrated how the first-order overlap $O(\rho_1,\rho_2)$ and the purity $\chi$ can be measured directly.

\section{Experimental considerations \label{sec:experiment}}

The setup in Fig.~\ref{fig:purity} provides a simple  way to
understand how the entire purity measurement protocol works.
It is, however, impractical from the experimental point of
view. First of all, it requires eight detectors that have to
be calibrated to give the same detection efficiency. Further,
to do that, it requires also six beam splitters (BSs) that have to
be well adjusted in a real experimental setup, and even if
this could be done, their number diminishes the effectiveness
of the protocol. The  setup, as depicted in
Fig.~\ref{fig:purity}, gives the maximum success probability
(conclusive coincidences) in 1/4 of the cases when all
photons are delivered. This value can be increased together
with experimental simplification of the procedure in several
ways. In this section we discuss several strategies
experimentally suitable for  various linear-optical
platforms.

\begin{figure}
\includegraphics[width=3.0cm]{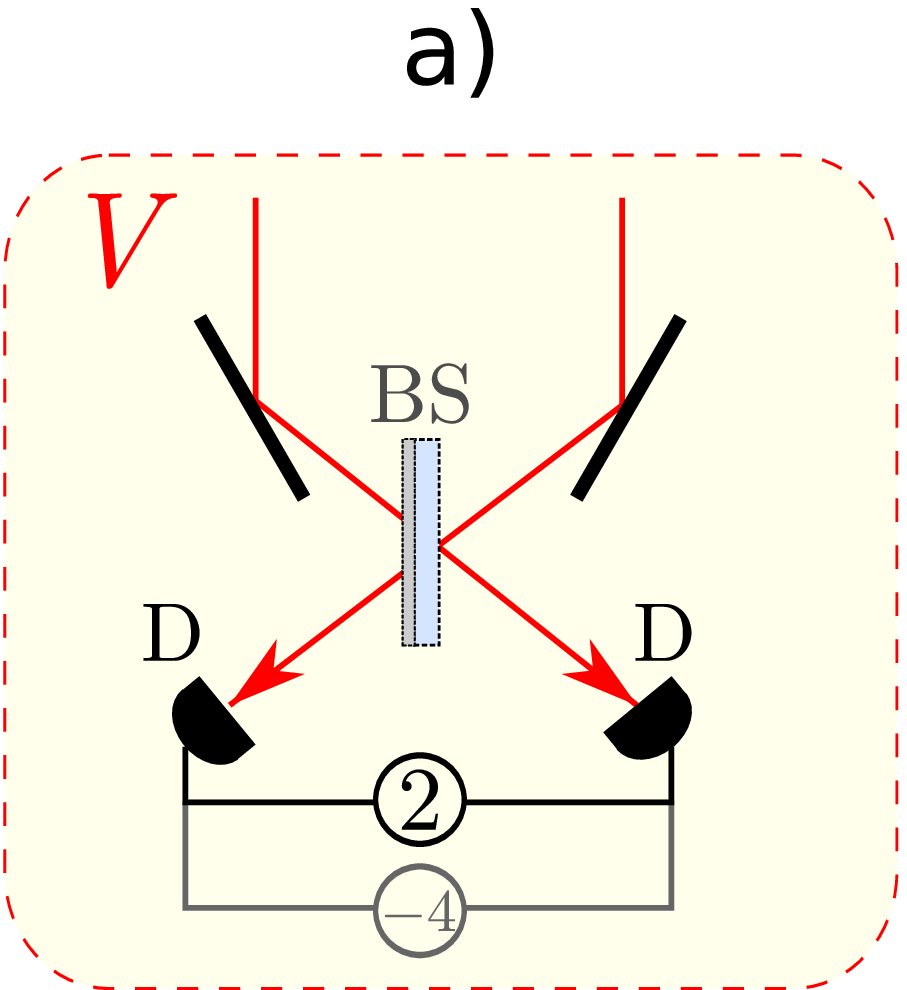} \hspace{2em}\includegraphics[width=3.0cm]{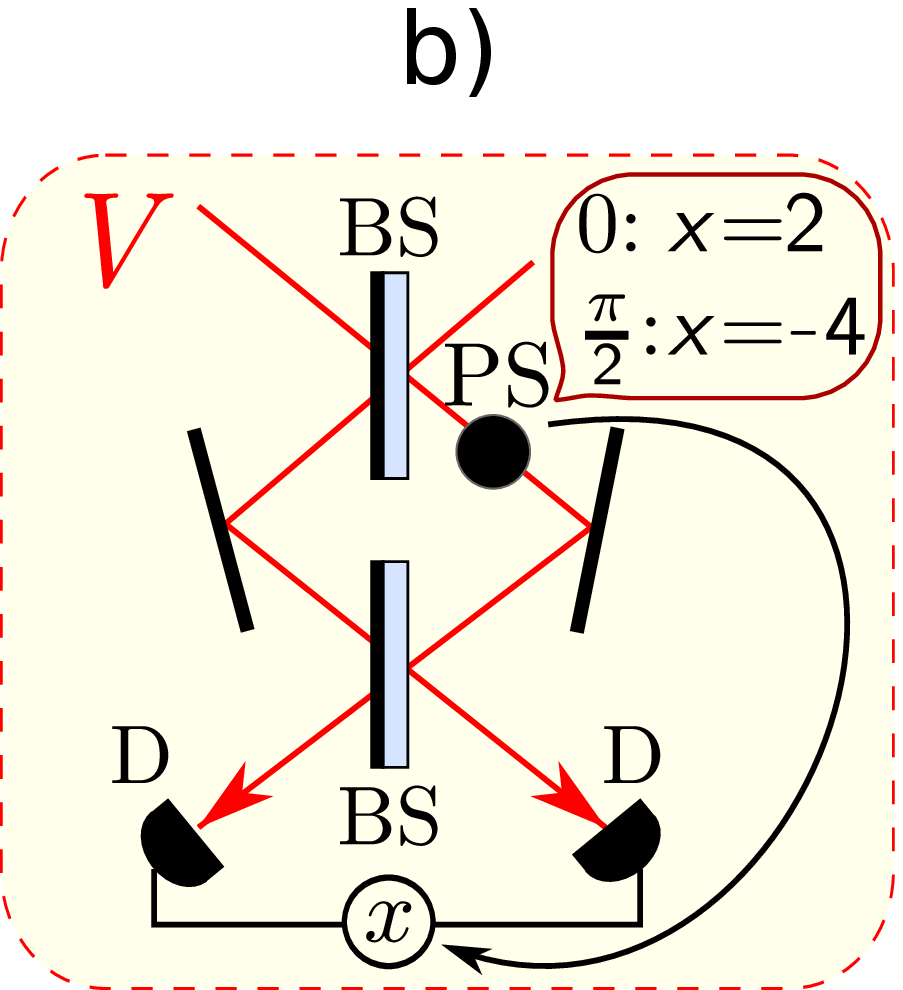}\vspace{2em}
\includegraphics[width=3.0cm]{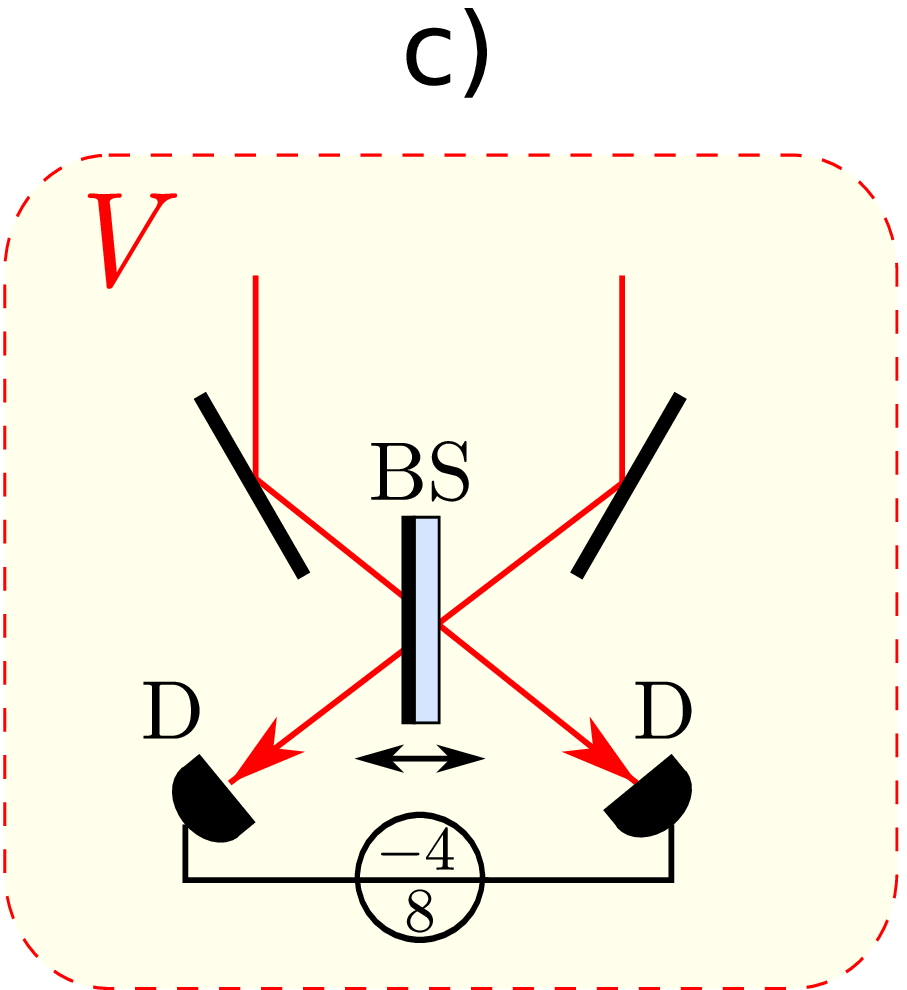} \hspace{2em}\includegraphics[width=3.0cm]{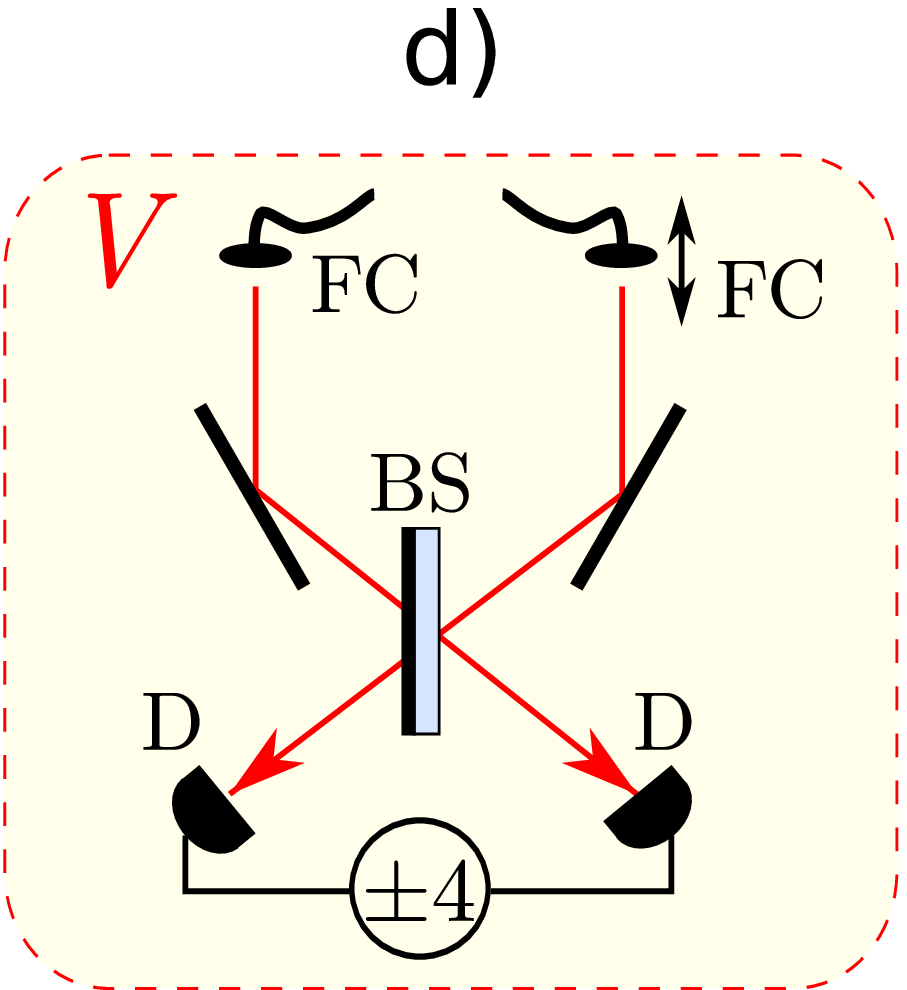}
\caption{\label{fig:experimental} (Color online)
Experimentally friendly methods for the measurement of the $V$
block as described in the text based on (a) a removable beam
splitter, (b) a Mach-Zehnder interferometer, (c) a partially movable
beam splitter, and (d) two-photon overlap alignment. BS,
balanced beam splitter, PS, phase shifter, FC, fiber coupler,
and D, detector. Motorized translation (double arrow) is used to
tune the temporal delay between the photons in order to switch
between the measurement regimes as explained in the text.}
\end{figure}

The first strategy involves just a removable beam splitter and a
pair of detectors [see Fig.~\ref{fig:experimental}(a)]. In this
case, one inserts a balanced beam splitter to the setup for a
half of the measurement time (assuming that photons are distributed uniformly in time). Thus, the singlet-state projection is performed
with the beam splitter inserted, while the identity projection with
the beam splitter removed. As in the original method, the
coincidence counts with the beam splitter are attributed to the value
$(-4)$, while those without the beam splitter are multiplied by
the value of 2. The effectiveness of this method, as defined in the
caption of Fig.~\ref{fig:purity}, reaches 1. Also the number of
beam splitters is reduced to two (one for each $V$ block), while the
number of detectors is reduced to  four. Unfortunately, this strategy would
not be very experimentally friendly in the case of bulk or
integrated optics. On both these platforms, removing and
reinserting the beam splitter is accompanied by a demanding
adjustment especially if the beam splitter has to be as balanced
as possible. In fiber-optics, this strategy can be implemented
more easily with  beam splitters with a tunable splitting
ratio~\cite{Bartuskova07}.

The second strategy is a generalization of the first strategy
bringing it closer to automation. It is depicted in
Fig.~\ref{fig:experimental}(b) and it involves using a
Mach-Zehnder interferometer that effectively implements a tunable
beam splitter. By changing the phase from 0 to $\pi/2$ in the interferometer, one
can achieve a splitting ratio in the range from 50:50 to 0:100.
This technique is particularly suitable for integrated
optics~\cite{Smith09}. In bulk optics and fiber-optics it is
impractical because of the experimental demands on keeping the
phase in the interferometer stable through the entire measurement.
Note that the number of beam splitters is reduced to four and so is
the number of detectors. The observed coincidences with the
interferometer set to the 50:50 ratio are multiplied by $(-4)$ and
those in the 0:100 regime are multiplied by 2. The maximum success
probability of this strategy is 1.

The third strategy is also a modification of the first one with
the exception that the beam splitter is not be removed completely,
but just shifted in one direction as depicted in Fig.
\ref{fig:experimental}(c). In our experience, this technique is
particularly useful in bulk optics. After shifting the beam
splitter out of its position, the reflected light no longer
couples to the detectors while the transmitted light still does.
Thus, we can effectively reach the splitting ratio 0:100, even
with a balanced beam splitter at the expense of some losses. The
readjustment of the beam splitter back to its balanced position is
easily done using a good quality  translation stage. As in
the first strategy, the number of detectors needed is four and the
number of beam splitters is two. Since only 1/4 of the impinging photons
reach the detectors if the beam splitter is shifted out (in),
we just multiply the number of such coincidences by 8 (-4) 
assuming that the measurements with the beam splitter in and
out take the same time. This technique has been used in several of
our experiments to measure technological losses in bulk
setups~\cite{Lemr11cpg,Lemr12clone}. It is, however, not suitable
for integrated optics and fiber optics. Its overall maximal success
probability is 5/8.

The last strategy to be discussed in this section is depicted
in Fig.~\ref{fig:experimental}(d). It involves a fixed
balanced beam splitter and a pair of detectors in each $V$
block. In order to implement singlet-state projection,
two-photon overlap, corresponding to 
Hong-Ou-Mandel interference~\cite{Hong87}, is obtained by
suitable position of the translation stage holding the output
couplers. On the other hand, the intensity projection is
implemented by deliberate misalignment of the output coupler
position so that the photons are separated in time by a
duration much longer than their coherence
time~\cite{Halenkova12appl}. Since the two-photon overlap has
to be adjusted anyway, a motorized translation stage is an
experimental necessity. With respect to that, this strategy
does not impose any new demands on the experimental setup.
The efficiency of this technique gives a success
probability of 3/4 since with the misaligned overlap, the
photons mark coincidence in only one-half of the cases.
Assuming the same measurement time with the output couplers
aligned and misaligned, the coincidences in the aligned position
are multiplied by $(-4),$ while those in the misaligned position
are multiplied by 4 in order to take into account the probabilistic nature
of such events.

\section{Efficient measurements of second-order overlap and
subfidelity \label{sec:setup2}}

Measuring the first-order overlaps $O(\rho_1,\rho_2)$,
$\chi(\rho_1)=O(\rho_1,\rho_1)$, and
$\chi(\rho_2)=O(\rho_2,\rho_2)$ is enough for determination of
the superfidelity $G(\rho_1,\rho_2)$.  If it is known that one of
the states is pure, then we do not need to proceed with estimating
the subfidelity because in this case we already have all the data
needed for calculating the fidelity
$F(\rho_1,\rho_2)=O(\rho_1,\rho_2)$.

In the simplest qubit case, the superfidelity and fidelity are
equivalent, i.e., $G(\rho_1,\rho_2)=F(\rho_1,\rho_2)$ and no
further work is required for estimating the fidelity. However, in
the case of quartits one also has to estimate the subfidelity
$E(\rho_1,\rho_2)$ to know in what range the fidelity
$F(\rho_1,\rho_2)$ falls. The only missing quantity needed for
estimating the subfidelity $E(\rho_1,\rho_2)$ is the second-order
overlap $O'(\rho_1,\rho_2)$, which depends on the Hilbert-space
dimension of a given system (i.e., qubit or quartit) and requires
from four to eight photons.

\begin{figure}

\fig{\includegraphics[width=7cm]{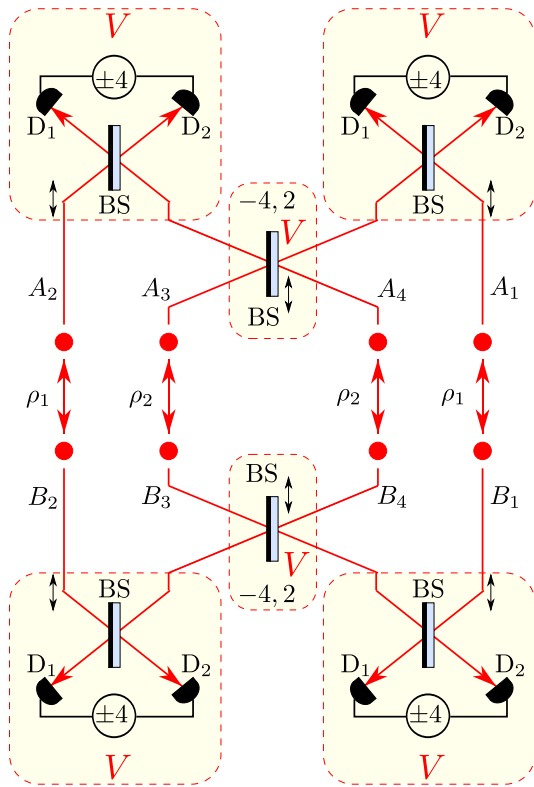}}

\caption{\label{fig:O2} (Color online) Proposal of the measurement
setup for the second-order overlap $O'(\rho_1,\rho_2)$ for
two-qubit states. The input is given as
$(\rho_1\kron\rho_1\kron\rho_2\kron\rho_2)_{A_1B_1A_2B_2A_3B_3A_4B_4}$.
If the states are single-qubit states then one of the blocks can
be removed. Note that in our optimized methods (as described in
Sec.~\ref{sec:experiment}), the number of required detectors is
equal to the number of photons. Double arrows indicate the
two-photon overlap alignment as in
Fig.~\ref{fig:experimental}(d).}
\end{figure}

So, our goal now is to describe a method for measuring the second-order overlap
$O'(\rho_1,\rho_2)$. We know that $ (\rho_1\rho_2)^2 =
\frac{1}{256} R^{(1)}_{ma}R^{(2)}_{nb}R^{(1)}_{kc}R^{(2)}_{ld}
(\sigma_m\sigma_n\sigma_k\sigma_l)\kron(\sigma_a\sigma_b
\sigma_c\sigma_d)$ and it can be verified that
\begin{eqnarray}
\mbox{Tr }(\sigma_m\sigma_n\sigma_k\sigma_l) = \mathcal{K}_{mnkl}\,,
\end{eqnarray}
where
\begin{eqnarray}
\tfrac{1}{2}\mathcal{K}_{mnkl}&=&
\kd{mn}\kd{kl} + \kd{nk}\kd{ml} - \kd{mk}\kd{nl}
+2\kd{m0}\kd{nl}\kd{mk}\nonumber \\ &&+ 2\kd{l0}\kd{nl}\kd{mk}
-4\kd{m0}\kd{n0}\kd{k0}\kd{l0} + \I\kd{m0}\e{nkl} \nonumber \\ &&+
\I\kd{n,0}\e{klm}+ \I\kd{k0}\e{lmn} + \I\kd{l0}\e{mnk}\ \label{N1}
\end{eqnarray}
is the kernel for which we describe the measurement method. So,
the investigated quantity reads
\begin{eqnarray}
\mbox{Tr }(\rho_1\rho_2)^2 &=&\frac{1}{256} \mathcal{K}_{mnkl}
\mathcal{K}_{abcd}\nonumber\\
&&\times\mbox{Tr }
(\gamma_{mnkl}\kron\gamma_{abcd})(\rho_1\kron\rho_2)^{\kron
2},\quad
\end{eqnarray}
where $\gamma_{mnkl}=
\sigma_m\kron\sigma_n\kron\sigma_k\kron\sigma_l$. In order to
design an efficient setup we have to calculate
\begin{equation}
H = \frac{1}{16}\mathcal{K}_{mnkl}\gamma_{mnkl},
\end{equation}
and find the most convenient permutation of the qubits comprising
the  eight-qubit system. This is because
\begin{eqnarray}
\mbox{Tr }(\rho_1\rho_2)^2  =  \mbox{Tr } [(H\kron
H)_{A_1A_2A_3A_4B_1B_2B_3B_4}\nonumber\\
\times(\rho_1\kron\rho_2\kron\rho_1\kron\rho_2)_{A_1B_1A_2B_2A_3B_3A_4B_4}].
\end{eqnarray}
The matrix $H$ is a permutation, so it can be decomposed into a
product of inversions, i.e., the SWAP operations. Note that the
$V$ matrix is equivalent to the SWAP operation (see also
Ref.~\cite{Ekert02}) as $V = 2S$. The result reads 
$(S_{34}S_{23}) H (S_{23}S_{34}) =\frac{1}{8} S_{23}(V_{12}\kron
V_{34})S_{23}V_{34}$. Thus, we have
\begin{eqnarray}
\mbox{Tr }(\rho_1\rho_2)^2  &=& \tfrac{1}{64} \mbox{Tr } \lbrace
[(V^{\kron 2})'(I\kron V)]^{\kron
2}(\rho_1\kron\rho_2\kron\rho_1\kron\rho_2)' \rbrace\nonumber \\
& =&\tfrac{1}{64} \mbox{Tr } \lbrace [(I\otimes V)(V^{\otimes 2})']^{\otimes 2}(\rho_1\otimes\rho_2\otimes\rho_1\otimes\rho_2)' \rbrace,\nonumber \\\label{Eq:32}
\end{eqnarray}
where $I$ in this section is the two-qubit identity operator,$(\rho_1\kron\rho_2\kron\rho_1\kron\rho_2)'
=(\rho_1\kron\rho_1\kron\rho_2\kron\rho_2)_{A_1B_1A_2B_2A_3B_3A_4B_4}
$ and the second-order overlap operator reads $[(V^{\kron
2})'(I\kron V)]^{\kron 2} = [(V^{\kron 2})_{A_1A_3A_2A_4}(I\kron
V)_{A_1A_2A_3A_4}]\kron[(V^{\kron 2})_{B_1B_3B_2B_4}(I\kron
V)_{B_1B_2B_3B_4}]$, which is a product of two Hermitian operators
that commute [see Eq.~(\ref{Eq:32})] for the input state
$(\rho_1^{\otimes 2}\otimes\rho_2^{\otimes2})_{A_1B_1A_2B_2A_3B_3A_4B_4}$,
thus one can measure these two operators subsequently. 
In the final step we just need to express
$I\kron V$ as $I\kron V = 2I^{\kron 2} - 4I\kron P^-$, where $P^-$
is the singlet-state projection that can be performed by a beam
splitter. Thus, as shown in Fig.~\ref{fig:O2}, we can perform the
measurement using the approach discussed in the previous section.
Here the beam splitter is placed  at or removed to the mixed modes $A_3$ and
$A_4$, or $B_3$ and $B_4$. For each of the four cases, the final
step consists of performing the measurement on four $V$ boxes.
Note that the problem of measuring the
second-order overlap is equivalent to the problem  of
measuring a first-order overlap of four-qubit states (i.e., a
16-level qudit or quantum hex).

From the experimentalist point of view, the setup presented
in this section is based on our fourth strategy as described in
Sec.~\ref{sec:experiment}. This particular choice of
implementation technique can be useful in bulk optics, where
motorized translation with a range on the order of centimeters is
often used to stabilize the two-photon overlap. On the other hand,
for  fiber and integrated optics implementations, this
technique might not be practical. However, it is straightforward to
swap the $V$ blocks in the proposed setup for any of the four
alternative implementations of the $V$ blocks proposed in
Sec.~\ref{sec:experiment}. The functionality of the setup remains
unaffected by this replacement.

\section{Conclusions \label{sec:conclusion}}

We proposed an effective and direct method for estimating the
bounds of the Uhlmann-Jozsa fidelity (or, equivalently, the
Bures distance) for two unknown mixed two-qubit states
without recourse to quantum state tomography. Namely, we
described how to measure the superfidelity and subfidelity,
which are the upper and lower bounds of the fidelity,
respectively~\cite{Miszczak09}. We showed explicitly that the
overlap of two density matrices is an observable which has
its own Hermitian operator $\Gamma$, given by
Eq.~(\ref{Gamma}), and furthermore can be directly 
measured in a linear-optical experiment. Our
method for the determination of the superfidelity is based on
the measurement of the first-order overlap. In a special
case, the method can be used for measuring the purity (or
linear entropy) of a mixed two-qubit state. On the other
hand, our proposal of a experimentally-friendly method for
direct measuring the second-order overlap of two arbitrary
two-qubit states enables the determination of their
subfidelity.

Concerning the purity measurement, it is worth referring
to Ref.~\cite{Hendrych03}, where its authors make use of 
Hong-Ou-Mandel interference and a singlet-state projection
performed by a balanced beam splitter to determine the
two-photon overlap. In a special case this technique can be
adapted for measuring the purity of a single qubit. A similar
setup was used in Ref.~\cite{Adamson08}, where the authors
described direct and indirect purity measurements of a single
qubit. They presented two separate strategies, one corresponding to a
direct measurement using many consecutive pairs of the
investigated state and the second corresponding to  the standard quantum
state tomography based on polarization projection
measurements. Since the authors of Ref.~\cite{Adamson08} were
dealing with the measurement of single qubits, their setup
requires fewer resources than the scheme presented in our
paper dedicated to two-qubit purity measurements.

Our $V$ block is in fact the Hong-Ou-Mandel dip known
since the famous experiment of Ref.~\cite{Hong87}. Indeed,
the Hong-Ou-Mandel two-photon interferometer has been used to
implement a projection on a singlet state, for instance, in
Ref.~\cite{Hendrych03}. On the other hand, with the
development of experimental techniques, the four alternative
implementations of the $V$ block suitable for various
platforms (including bulk, fiber, and integrated optics) have
their merit.

As mentioned before, a few experiments on single-qubit
purity have already been performed but, since there are additional
interesting features (including entanglement) arising from the
transition from one to two qubits, our proposals for direct
measurement of the two-qubit purity can be an important tool for future
investigations in quantum optical engineering and information
processing.

Concerning the two-photon overlap measurements, we note
another formal method of Ekert \etal~ \cite{Ekert02} based on
programmable quantum networks with controlled-SWAP gates for
estimating both linear and nonlinear functionals of arbitrary
states. As shown explicitly by Miszczak \etal~ \cite{Miszczak09},
this network method also enables the measurement, in a special case, of the
first- and second-order overlaps between a pair of two-qubit
states for the estimation of their fidelity bounds. This formal
method, although in principle scalable for any number of qubits,
has not been applied (even theoretically) to any physical system.
In contrast, we apply a purely algebraic method for estimating
some specific functionals of states and describe an
experimentally-friendly linear-optical implementation of our
method. As in these related works~\cite{Ekert02,Miszczak09}, we
assumed that we have access to two copies of a given quantum
state, which can be implemented either by producing two identical
states simultaneously, or by storing the state produced earlier in
order to measure it together with the second copy of the state
available later. Concerning the optical measurement of the purity,
our proposal requires fewer experimental resources than of
Bovino \etal~\cite{Bovino05}. Specifically, our setup requires
only four instead of six detectors and only two beam splitters
instead of four used in Ref.~\cite{Bovino05}. Every additional
experimental resource complicates the operation of the setup and
is a possible source of imperfections.

In Sec.~\ref{sec:experiment} we have presented several
experimentally-friendly strategies based on our original method.
All these setups represent a viable alternative to be considered
in practical experimental implementations of the protocol. We
cannot claim that one of them is superior to the others since each
of them has its advantages and drawbacks. The choice would depend
on the preferred features. For instance, if the success rate is an
issue, experimentalists would probably choose the second strategy
(with a success rate up to 1). On the other hand, for the
highest experimental stability and precision (in bulk optics,
rather than in fiber optics), the fourth strategy might be
selected. Moreover, the third strategy would be a good choice for
a fiber-optical implementation, while the second strategy can be
suitable for integrated optics.

Moreover, we performed a Monte Carlo simulation of $10^7$ mixed
two-qubit states and applied the method of least squares to
estimate the fidelity as a generalized power mean of these
fidelity bounds with the minimum average estimation error. 

Experimentalists frequently use the fidelity (usually
determined by applying quantum tomography) as a measure of
the quality of their achievements. Remarkable progress has been
observed over the past decade as the reported experimental
values increased from 0.58 to 0.98.

However, it should be stressed that both the purity and
fidelity are important parameters for characterizing
experimental achievements with respect to quantum operations.
The purity alone would not be adequate to describe how well
the gate performs the requested task (for instance the controlled NOT
operation). Similarly to that, the fidelity can be indeed rather
high even though the purity has dropped and therefore the
gate is misaligned, which typically occurs in the case of
active stabilization issues. As pointed out recently in
Ref.~\cite{Dodonov12} to have 98\% fidelity is not enough to
be sure that the two states are indeed similar or close to
each other. The answer depends on the specific situation and
additional information about the state is needed.

We hope that this paper can stimulate further
experimental interest in determining both the fidelity and
purity for the purposes of quantum engineering and quantum
information processing.

\begin{acknowledgments}
We thank Pawe\l{} Horodecki for inspiring discussions. The authors
acknowledge support by the Operational Program Research and
Development for Innovations -- European Regional Development Fund
(Project No. CZ.1.05/2.1.00/03.0058) and the Operational Program
Education for Competitiveness - European Social Fund (Projects No.
CZ.1.07/2.3.00/20.0017, No. CZ.1.07/2.3.00/20.0058, 
and No. CZ.1.07/2.3.00/30.0041) of the
Ministry of Education, Youth and Sports of the Czech Republic.
A.M. was supported by Grant No. DEC-2011/03/B/ST2/01903 of the
Polish National Science Center. K.L. also acknowledges support
by the Czech Science Foundation (Grant No. 13-31000P).
\end{acknowledgments}

\end{document}